\documentclass[journal]{IEEEtran}

\usepackage{graphicx}

\usepackage{array}
\usepackage{multirow}
\usepackage{longtable}
\usepackage{rotating}
\usepackage{multirow}
\usepackage{url}
\usepackage{textcomp}
\usepackage[T1]{fontenc}

\newcommand{\tabincell}[2]{\begin{tabular}{@{}#1@{}}#2\end{tabular}} 

\ifCLASSOPTIONcompsoc 
\usepackage[caption=false,font=normalsize,labelfont=sf,textfont=sf]{subfig}
\else
\usepackage[caption=false,font=footnotesize]{subfig}
\fi
\ifCLASSINFOpdf
\else
\fi
\begin{document}
	%
	
	\title{A Review on Serious Games for Disaster Relief}
	%
	%
	%
	\author{Huansheng Ning, Zhangfeng Pi, Wenxi Wang, Fadi Farha, Shunkun Yang
	\thanks{Huansheng Ning, Zhangfeng Pi, Wenxi Wang and Fadi Farha
are with the University of Science and Technology Beijing, Beijing
100083, China} 
	\thanks{Huansheng Ning is also with Beijing Engineering Research Center for Cyberspace Data Analysis and Applications, 100083, Beijing, China}
	\thanks{Shunkun Yang is with the Beihang University, Beijing}	
}

	%
	%

	\markboth{Journal of \LaTeX\ Class Files,~Vol.~14, No.~8, August~2020}%
	{Shell \MakeLowercase{\textit{et al.}}: A Review on Serious Games for Disaster Relief}
	%



	\maketitle
	
	\begin{abstract}
	Human beings have been affected by disasters from the beginning of life, bringing them many sad memories. In the long struggle against disaster, people have devised a variety of methods to train relevant participants in disaster relief capabilities. However, many traditional training methods, such as disaster exercises may not provide effective training to meet the need of today. Serious games provide an innovative approach to train participants in disaster relief, and a large number of Serious Games for Disaster Relief (SGDRs) have been developed to train disaster planning and rescue capabilities. At the same time, there is no systematics phase description for disaster relief, which cannot effectively guide participants' work and training in disaster relief. Therefore, this paper proposes a comprehensive and professional disaster relief classification framework according to different relief work in each stage of the disaster. Based on this framework, we review the functions and technologies of serious games in each classification, which can offer reliable guidance for researchers to better understand and use SGDRs. In addition, we analyze the serious games in each category, point out the limitations, and provide some valuable advice for developers on game design.
		
	\end{abstract}
	

	%
	\IEEEpeerreviewmaketitle

	\section{Introduction}
	%
	%
	%
	%
	\IEEEPARstart{D}{}isasters are catastrophic events that may overwhelm the emergency response capabilities of the community and threaten public safety and the environment \cite{1}. Disasters always have significant impact on societies, economies, and humankind. According to the Annual Disaster Statistical Review 2017, there were $335$ natural disasters that affected more than 95.6 million people, causing 9697 deaths and total losses of \$$335$ billion \cite{2}. Besides natural disasters, there are also many man-made disasters, such as fire, terrorist attacks, transport accidents, and technological disasters, which cause considerable victims and economic losses \cite{3}. These disasters will bring people a great physical and mental pain, even deprive a number of lives. Therefore, it is vital to train relevant participants in disaster relief capabilities. Generally, the disaster relief capability training adopts traditional approaches such as videos, posters, disaster exercises, etc. However, the major limitation of these approaches is that they cannot fully produce the elements of the disaster, which cannot ensure that participants receive effective training. Serious game, with non-entertainment purposes, is an innovative approach which can create an immersed disaster environment by using game elements, e.g. symbolic tokens, models and sound to simulate the impact of a disaster, e.g. destroyed building, injured people \cite{4}. Connolly \cite{5} has shown that participants in disaster relief can obtain more effective training by serious games compared with traditional approaches. Therefore, serious games have taken its place in the disaster relief training.
	
	Hence, serious games for disaster relief training have drawn widespread attention, and many related researches have been done. But most studies focused on the certain stage of a disaster and did not cover the entire disaster process. However, a successful disaster relief process should involve all possible activities and situations at all stages of a disaster \cite{6}. Disaster relief training needs to train different capabilities in each stage of a disaster. Therefore, it is necessary to classify the relief work in different stages of a disaster. Disaster management \cite{6} is defined as the organization and management of resources and responsibilities for all stages of a disaster, which can effectively guide disaster relief actions in each stage of a disaster. Because there has not been a systematic model for all stages of a disaster relief yet, we integrate the idea of the disaster management into the entire disaster relief process. According to the International Federation of Red Cross and Red Crescent Societies (IFRC), disaster relief is divided into three stages: Reparedness, Response, and Recovery \cite{7}. By contrast, most literature and organizations believe that disaster relief consists of four phases base on disaster management-Mitigation, Preparedness, Response, and Recovery \cite{8}. We analyze these two frameworks and conclude a new framework to describe the entire disaster relief process. Then, we reviewed a quantity of Serious Games for Disaster Relief (SGDRs) in our framework and analyze their characteristics, target group, techniques, and possible disaster relief capabilities so as to offer reliable guidance for relevant participants, e.g. disaster communities, rescuers, policymakers and incident commanders.
	
	This paper discusses the limitations of SGDRs from multiple aspects, and further proposes some suggestions for developers to design more effective serious games. The rest of the paper is organized as follows: section 2 introduces traditional training methods as well as serious games training methods, and then presents the stage classification disaster relief framework according to the different tasks in each stage. Section 3 surveys the SGDRs in each disaster relief stage. Section 4 analyzes the deficiencies of SGDRs and proposes the corresponding countermeasures. Finally, section 5. draws the conclusions and future direction.
	

	\section{Background}

	
	\subsection{Disaster Relief Training Methods}
	Disaster relief is a fundamental process to fight disasters. Effective disaster relief can greatly reduce the losses caused by disasters. In disaster relief, relief workers have to work under a great pressure to make decisions and implement an appropriate action. In contrast, the training aim to provide individual with the knowledge, skills, and attitudes to cope with potential stressors \cite{9}. Consequently, effective disaster exercise can greatly improve participants’ ability to deal with disasters. In general, people often choose to use operation-based methods such as disaster drills to practice and maintain rescue capabilities and use discussion-based methods such as tabletop exercise to develop and assess plans, policies, and procedures.

	\begin{table*}
		\centering
		\caption{The Main Approaches for Disaster Relief training}
		\label{table}
		\setlength{\tabcolsep}{3pt} 
		\renewcommand\arraystretch{1.5} 
		\begin{tabular}{|m{3.0cm}<{\centering}|m{3.0cm}<{\centering}|m{6.0cm}<{\centering}|m{6.0cm}<{\centering}|}
			\hline 	
			Approaches&Target abilities&Advantages&Limitation \\ 
			\hline
			Disaster drill & Rescue skills &\tabincell{c}{ benefit to train team collaboration,\\improve and evaluate \\ local disaster response capacity, etc.} & \tabincell{c}{expensive,\\ unfavorable to newcomer, \\different from real disaster rescue,\\rescuer maybe injury in training, etc.}\\
			\hline
			Tabletop exercise&Disaster planning&\tabincell{c}{low stress environment,\\low cost,gathering ideas and wisdom,\\facilitated group discussion of problem, etc.}&\tabincell{c}{difficult to organize,\\lack of realism,\\unable to replicate every aspect\\ of a hypothetical situation,\\ providing only a superficial review, etc.}\\
			\hline
			Serious game&\tabincell{c}{All disaster relief\\ capabilities}&\tabincell{c}{safe,
				low cost,\\
				customizable scenes,\\
				repeatable training,\\
				simulating scenes that are difficult\\ to reproduce in reality, etc.}
			&\tabincell{c}{the discomfort of game\\ technology to the human body,\\
				people do not take it seriously,\\
				ignoring the mistakes in the game,\\
				inability to fully express the\\ complexity of the disaster, etc.\\
			}\\
			\hline	
		\end{tabular}
		\label{tab1}
	\end{table*}

	It is generally know that disaster drills are coordinated, supervised activities, usually used to test specific operations or functions \cite{10}. What’s more, the disaster drill training method is widely used around the world and is regarded as a fundamental tool for evaluating and improving local disaster response capacity \cite{11}. In the disaster drill, Rescuers in different types of training scenarios affected by the disaster where victims are replaced with dummies, rescuer needs to use various rescue skills to keep “victims” safe. Therefore, the disaster drill is effective in improving personal disaster rescue capabilities, but not optimal for all situations. On the one hand, it is relatively expensive because it requires items to be consumed and enter a dedicated training area, where various environments (such as buildings, ships, trains, etc.) need to be built. On the other hand, the drill may be overwhelming to the newcomer, especially when involving large scale simulations \cite{12}. In fact, rescue behavior in real life is different from experiments such as drills \cite{13}. Additionally, rescuers are likely to be injured during the drill \cite{14}, which means that current drill training still has limitations.
	
	Table exercise is a discussion-based learning experience where participants need to play a role and use their strategies to solve problems \cite{15}. During the discussion, participants can not only strengthen communication, but also enable to evaluate the effectiveness of emergency response strategies. This method often used to train disaster planning capabilities, such as, the strategies to reduce disaster risk, assess plans and policies about disasters. For example, Taylor et al. use tabletop exercise to train officials’ strategies to fight Pandemic Influenza \cite{16}.  H Khankeh et al. use tabletop exercise to train hospital managers to plan for disaster \cite{17}. AM Wendelboe et al. use tabletop exercise to review and test the measures taken by university leaders to deal with the COVID-19 \cite{18}. These experiments are enough to show that tabletop exercises are necessary for disaster relief planning and future disaster prevention. However, tabletop exercises also have their limitations. It is difficult to satisfy the requirement of multiplayer participation where  officials, policymakers, managers, and experts must gather in one place. Apart from that, same as the disaster drillt it lacks reality, different from the real situation, and difficult to consider all aspects, provide only a superficial review of the overall plan \cite{19}. Consequently, it is necessary to investigate innovative and more effective approaches to overcome these limitations.
	
	Considering the apparent limitations of traditional approaches in disaster relief training, serious games have been used as an alternative for training instead of traditional exercise. Serious games are using game technologies and game elements for applications that aim to learn or train, not only for entertainment \cite{20}. Unlike traditional disaster drills and tabletop exercises, serious games can easily simulate the element of disaster by using game elements, such as symbolic tokens, models, sound effects, virtual reality, and so on. In this context, each participant may experience a certain key characteristic of a real disaster that enables them to better understand the disaster. Besides, there are many other benefits to the serious games training method based on the characteristics of the game. It allows more frequent training, possibilities to train that are not easy to reproduce in the real world due to cost, safety, and time concerns, and allows better evaluation \cite{21}. At the same time, it provides a safe, low-cost alternative that can be practiced in certain situations, and provides trainees with the opportunity to train various work-related procedures. With the rising popularity of various types of games such as electronic games, VR games, and somatosensory games continue to increase the upper limit of serious game training. While enjoying the high immersion brought by these technologies, we also need to overcome their shortcomings. Such as eyesight problems caused by electronic devices, dizziness and sickness caused by VR technology \cite{22}, the discomfort caused by somatosensory clothing, and so on. This requires us reasonably plan the time for training using these techniques. In addition, serious games as a type of game also have some limitations. Firstly, serious games maybe build up a false sense of security \cite{21}, because in serious games, players are free to make mistakes without any actual punishment, especially in games without a good feedback mechanism, which make players ignore these mistakes even in the real world. Secondly, certain groups are not familiar with what serious games are or the difference between serious games and games. Considering games made it certain that users did not take serious games seriously. Finally, serious games may inevitably simplify disaster, and thus fail to adequately portray the complexity of the disaster and the process of disaster relief. For example, terrain control of lava flow and tsunami distribution, the relationship between earthquake intensity and source distance \cite{23}, as well as excessive use of water in a fire may create a steam cloud that will pose a major threat to firefighters and any potential victim being rescued. These details are often difficult to consider in the game \cite{24}. 
	
	Therefore, serious games cannot completely replace traditional disaster relief training, but taking serious game training methods as a supplement to traditional training methods can reduce training costs and provide a specific environment for training specific tasks. At the same time, serious games can also be used as a prelude to traditional training. After the trainees use serious games to reach a certain level, they contact with expensive and resource-intensive traditional training to ensure the efficiency of training.

	\subsection{Disaster Relief Cycle}
	Disaster relief is a complex process that usually requires a large number of measures to deal with a series of uncertain emergencies. That normally requires various actors (e.g. civilians, government, and non-government) to perform their duties and coordinate with each other at each stage of the disaster. At the same time, serious games have been widely used in relief training for different groups of people at different stages of disasters. Therefore, it is very necessary to distinguish relief work in different stages of the disaster and select the appropriate SGDRs to train different abilities. Now there are two main frameworks that describe the stage of disaster relief based on disaster management. The first framework is proposed by The Federal Emergency Management Agency (FEMA) \cite{25} to divide this process into four phases, while the second format is proposed by IFRC \cite{7} to divide the process into three phases. Both frameworks can clearly express the relief work at each stage of the disaster. However, their descriptions of the disaster relief process have their own characteristic. Based on this, we further analyzed the characteristics of these two frameworks and concluded a comprehensive and professional disaster relief classification framework. The framework is not only used as a guideline to guide participants in disaster relief to take appropriate relief work at different stages of the disaster but also can as a classification standard for SGDRs.
	\begin{figure}[h]
		\centering
		\includegraphics[width=3.7in]{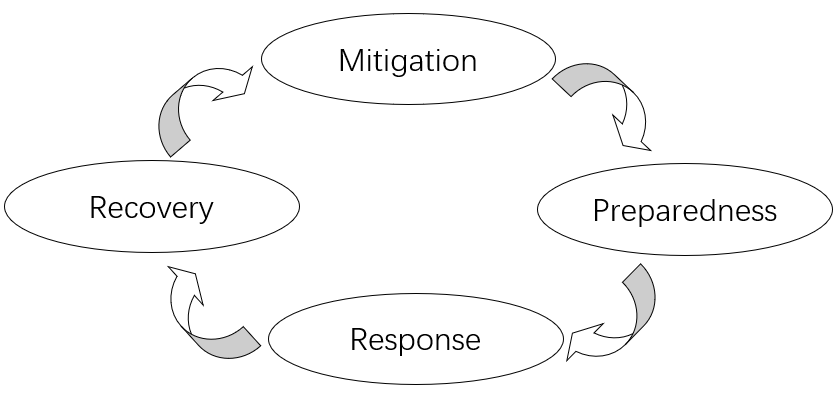}
		\caption{The Framework of FEMA}
	\end{figure}
	
	The FEMA \cite{25} proposed a basic four-stage framework, which is also widely used by most articles and organizations. As show in Fig. 1. It includes Mitigation, Preparedness, Response, and Recovery. It cycles through each stage and takes different relevant relief work in different phase. For example, disaster mitigation is the stage to eliminate or reduce the probability of disaster occurrence or reduce the negative effects of inevitable disasters. People take activities such as disaster analyses, disaster forecasting, and disaster defense project. Disaster preparedness refers to increase the likelihood of successful disaster response, such as taking a disaster response plan and raising public awareness would be considered preparedness activities. The purpose of disaster response is to respond to a disaster as rapidly as possible, by mobilizing resources to rescue survivors and meet their basic needs. Disaster recovery aim to assist those who have suffered the impact of a disaster to return the normal life. Both the mitigation stage and the preparedness stage occur before the disaster, and the disaster preparedness stage is complementary to the disaster mitigation stage, resulting in the relief work in these two phases is always closely linked. For example, people always carry out disaster prediction and disaster planning at the same time to ensure a timely response to the disaster. Similarly, most of the related serious games always pair preparedness skills with mitigation knowledge. Therefore, combining these two stages can more concisely and clearly show the relief work in pre-disaster. 
	
	\begin{figure}[h]
		\centering
		\includegraphics[width=3.65in]{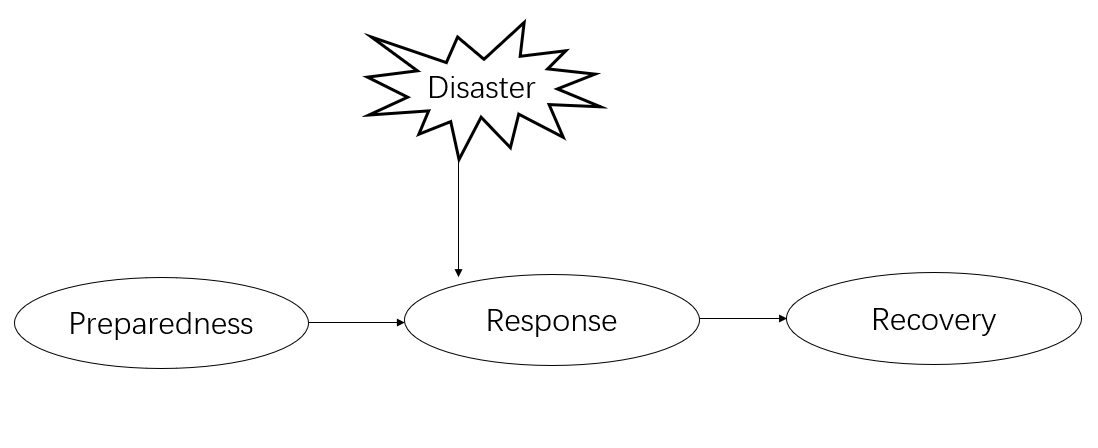}
		\caption{The Framework of IFRC}
	\end{figure}

	To address this issue, we analyzed the IFRC’s \cite{7} disaster relief phase division. IFRC divides the disaster relief process into three stages based on disaster management—Preparedness, Response, and Recovery. As showed in Fig.2. Unlike FEMA, IFRC combines the relief work before the disaster into one stage to express the disaster relief process more simply and clearly. In the preparedness stage, people take activities that provide relief measures to reduce potential disaster areas vulnerability to disasters and strengthen their capacities to respond to disasters. Another difference is that the framework depicts the phases of disaster relief as a linear process. But the disaster relief phase can best be represented as a cycle which is very important to the disaster relief process. Because disaster does not just appear one day, post-disaster review always is carried out in the recovery period. Such review will often reveal the shortcoming in the previous disaster plan, and then provides valuable experience and strategies for subsequent disaster preparedness \cite{6}. 
	
	\begin{figure}[h]
		\centering
		\includegraphics[width= 3.5in]{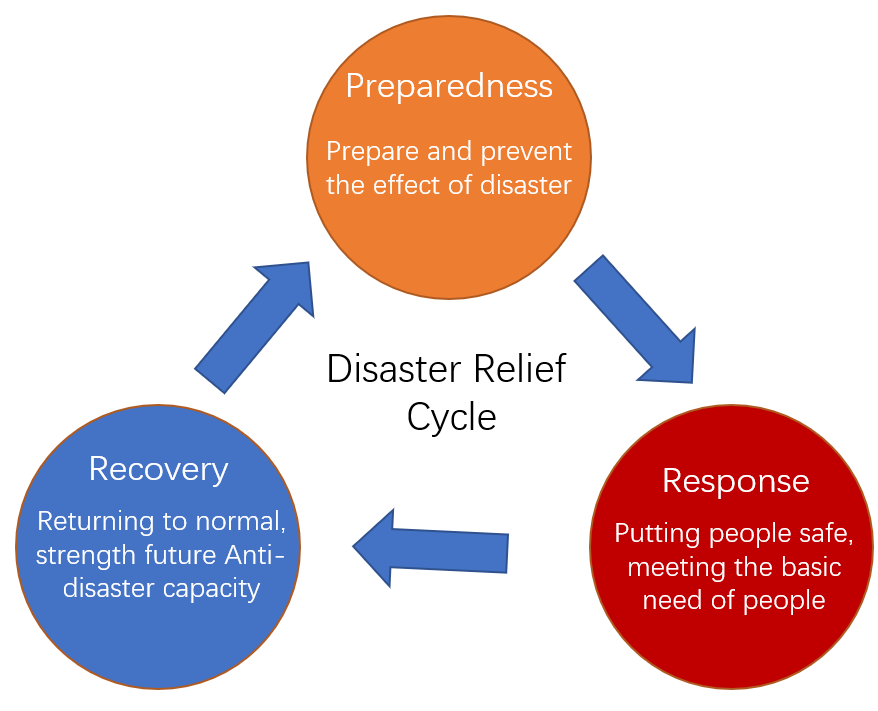}
		\caption{Disaster Relief Cycle}
	\end{figure}

	Therefore, we combined the above two frameworks to design a more comprehensive framework as a standard for dividing the stages of the disaster relief cycle. As show in Fig.3. On one hand, we divided the disaster relief process into three stages, namely Preparedness, Response, and Recovery. Each of phase levies particular demands on participants in disaster relief. In the preparedness stage, adopt policies, take a disaster response plan, etc. to prevent the impact of the disaster. In the response stage, some rescue actions are taken to keep people safe, and emergency supplies are provided to the disaster area to ensure the basic needs of the refugee. In the recovery stage, the need to help the reconstruction of the disaster area and strengthen the community’s anti-disaster capacity. On the other hand, these phases follow one another in a continuous cycle, emphasizing the positive impact of previous experience on current planning. In summary, SGDRs of each stage should also exhibit these properties. Later, based on this disaster relief framework, SGDRs at each stage of the disaster will be reviewed.

	\begin{table*}
		\centering
		\caption{Relief Works at Each Stages of Disaster Relief}
		\label{table}
		\setlength{\tabcolsep}{3pt} 
		\renewcommand\arraystretch{1.5} 
		\begin{tabular}{|m{3.0cm}<{\centering}|m{3.0cm}<{\centering}|m{6.0cm}<{\centering}|m{6.0cm}<{\centering}|}
			\hline 	
			Stage&Time&Characteristics&Relief Works \\ 
			\hline
			Preparedness & Pre-disaster &\tabincell{c}{ Prepare and prevent the effect of \\disaster} & \tabincell{c}{deploy disaster prevent projects,\\
				disaster policy and plan,\\
				raise public awareness,\\
				improve the environment, etc.\\
			}\\
			\hline
			Response&\tabincell{c}{During the disaster\\ and one or six months \\after}&\tabincell{c}{Putting people safe and meeting the\\ basic need of people}&\tabincell{c}{rescue victims in disaster,\\
				evacuate people,\\
				provide emergency supplies, etc.\\
			}\\
			\hline
			Recovery&\tabincell{c}{Six months after the \\disaster and lasts for\\ a long time}&\tabincell{c}{Recovery of infrastructure and\\ strength future Anti-disaster capacity.}
			&\tabincell{c}{rebuilding homes after the disaster,\\
				ensuring medical and health care in\\ disaster area, etc.
				
			}\\
			\hline	
		\end{tabular}
		\label{tab1}
	\end{table*}

	
	%
	
	\section{Serious Games for Different Stages in Disaster Relief}
	In this part, SGDRs are classified based on the framework mentioned above. As showed in TABLE II, according to different relief work at different stages of the disaster, the classification of SGDRs and related games are presented.

	\subsection{SGDRs for Preparedness Stage}
	The preparedness phase of disaster relief happens before the disaster or after the disaster recovery. This phase involves relief actions taken before an emergency to ensure a more effective response and steps to minimize the damage caused by a disaster, such as arranging the disaster response plans, constructing of disaster prevention projects, improving the environment, and increasing the skills and knowledge among the staff and community. Without sufficient disaster preparedness, people will be in a hurry when a disaster occurs, which causes huge casualties and property losses. Therefore, a number of serious games are designed for disaster preparedness, which allows people to understand disaster preparedness, especially that some games may train managers to predict, plan, and manage disasters before they occur.
	
	Many games depict on preparedness knowledge or enhancing civilian disaster awareness or evacuation skills. The knowledge conveyed by these games includes what to do, where to go before a potential disaster happens. Mannsverk \cite{26} and his team developed serious games to raise civilian awareness about the importance of preparation in facing floods. In Disaster Master \cite{27}, people learn how to identify the first signs of disasters and how to shelter during a disaster. The primary audience of Earth Girls \cite{28} is preteens which aim to help players better understand natural disasters through imaginative and interesting games. Jacob et al. \cite{29} developed a game called Smart Fire Safety, in which players can understand the fire safety hazards and precautions in the kitchen or gas station.
	
	In addition, there are many games that can improve the evacuation skills (e.g. system perception, pre-movement behavior, finding the route, exit choice, and navigation interactions). Rahouti et al. \cite{30} developed a game about fire fighting and evacuation training in medical institutions, which simulates a specific fire emergency and aims to train medical staff so that they will be able to provide evacuation instructions to the patient. This game is different from other systems because it can generate some 3D virtual patients with limited mobility, and the medical staff is expected to provide appropriate assistance to these virtual patients. In order to simulate the impact of crowd behavior on individual evacuation, some games use artificial intelligence technology to control the crowd or Non-Player Characters (NPCs) behavior. Ribeiro et al. \cite{31} applied crowd representation models (e.g. cellular automata models, forces-based models and, artificial intelligence-based Models) to represent the movements and behavior of crowed, and use human behavior model into the game to simulate human behavior factors in emergencies, such as panic, disorganized, irrational, etc. In this game, virtual crowds with different reactions and pedestrians with different behaviors will be generated. Players must overcome their instinct to follow other individuals and choose the most appropriate escape route based on their own judgment. The evaluation preliminary results show that this game has quite promised to train players’ judgment in disaster. In another study Ruffino et al. \cite{32} bring Building Information Modeling (BIM) and serious game together. This combination has great potential because using BIM to build models make them closely related to the real built world \cite{33}. This game is created based on a real building, and the game is divided into 4 levels. The higher the level, the greater the difficulty. Players must find the shortest path to the safety exit and run to the safety exit as fast as possible when a fire occurs. 
	
	Nowadays, VR technologies and augmented reality(AR) technologies have proved to be a valuable alternative to disaster evacuation training. Liang et al. \cite{34} developed a VR game to improve earthquake evacuation skills. In this game, the immersion of VR technology is used to simulate building shaking. Players can perceive earthquakes by observing the movement of buildings, ceiling lamps, and furniture in a virtual world. That provides players with knowledge on how to perceive earthquakes and respond safely. Unlike VR, which immerses users in a computer-generated environment, the more novel AR technology can combine the real world with virtual digital content, making training closer to reality \cite{35}. Catal et al. \cite{36} developed an AR-based evacuation system, which is suitable for fire, earthquake, and chemical attack. They implemented a GPS system to determine the player’s location and provided virtual instructions to players according to their location. Players had to follow the instructions to leave the building. Finally, they showed through experiments and questionnaires that the system was evaluated effectively, and most people could learn about disaster evacuation knowledge and skills.
	
	After the game development is completed, some authors \cite{31} \cite{32} have made game evaluations by different groups of people. They tested multiple players (half of them are familiar with route of the building in the game, and the other half are not familiar with the route). The result showed that players who are familiar with the building route had a shorter time to pass the game and were more efficient in mastering disaster evacuation skills. Therefore, for more effective training, these such serious games are not suitable be used by visitors or people who are not familiar with the building routes. For newcomers, before training by the serious game, they should be familiar with the route and environment of the building.
	
	The aforementioned games can raise public awareness and knowledge of disaster evacuation, but they are useless for training mangers’ disaster preparedness capabilities. There are also many SGDRs that are used for the disaster preparedness measures before the disaster. These games are mainly strategy role-playing games based on PC or website, developed by some simple game engines or web technologies. They often allow player to control emergency planning, disaster policies or city management. Players must choose from multiple options, and may observe the negative or positive consequences of their decisions when the disaster finally occurs. Floodsim \cite{37}, a simulation-based serious game, is developed by Mendez et al. In this game, the player is a policymaker in control of flood policy in the UK to minimize the impact of flooding on people. The player needs to make some policies according to population density, economic output, and flood risk. After that the game in the form of a newspaper presents results of taken decisions. Apparently, this game can easily verify the reasonable of the disaster policy. Apart from that, Stop Disaster \cite{38} is another disaster simulation strategy game developed by flash, which simulates five kinds of disasters to let players deploy certain resources or construct some man-made protections to protect the town from disasters within budget. The player's final score is closely related to the effectiveness of resource allocation.
	
	The process of disaster relief best is a cyclical process. The disaster preparedness and prevention plan before the disaster should be based on the experience of the previous disaster. For example, the strategy and plan for the fight against SARS in 2004 can provide valuable experience and guidance for the prevention of the new coronavirus in 2020. However, most of SGDRs for the preparedness stages almost no have emphasis on the importance of the experience of the previous disaster. Therefore, serious game designers need to take this into consideration when developing games in the future.

	\subsection{SGDRs for Response Stage}
	The response phase of disaster relief happens during the disaster and one or six months after. In this stage, the disaster has occurred and has taken a great impact on people. People take a large and complex set of activities that work to minimize the damage caused by the disaster and protect life or property. The primary aim of the response stage is to rescue victims from immediate danger. This requires the rescuer to take a series of rescue actions to search and rescue the victims. These actions are closely related to the command and plan of the incident commander. Therefore, SGDRs for the response stage is mainly divided into two types: one is that the player is a rescuer involved in a specific rescue mission, and the other is that the player takes the role of the incident commander to direct the disaster relief team. At the same time, it is also necessary to provide immediate relief (e.g. medical care, food and water, and temporary shelter, etc.) for the refugee to meet their basic needs. There are a few non-electronic games about providing capital or medical care to disaster areas. For example, Buzz about Dengue \cite{39} is a team-based strategy game that can teach players how to prevent Dengue Fever by providing medical health. Dissolving Disasters \cite{40} designed by Red Cross can help people understand the importance of donors to the refugees in disaster areas.
	
	\subsubsection{SGDRs for Rescuers}
	when a disaster occurs, the rescue workers (e.g. firefighter, soldier, medical staff, volunteer, etc.) are always on the first line of disaster response. While rescuers are saving people, their lives will also be threatened. According to the International Association of Fire Fighters (IAFF), the fire department have four times the rate of work-related injuries as the private industry. One in every three firefighters is injured while on their duties \cite{41}. Thus, in order to reduce the casualty rate of rescuers and improve the rescue efficiency of rescuers, a large of serious games have been developed for rescuers.
	
	Hazmat: Hotzone \cite{42} is an instructor-based simulation based on video game technology, developed by the Entertainment Technology Center at Carnegie Mellon University and the Fire Department of New York to train first responders to handle chemical and hazardous materials emergencies. In this game, the instructor has full control of any aspect of the simulation, such as specify the wind, temperature, precipitation for outside environments, the type and location of hazardous material, and where the victims are located. The instructor also can freely add some new elements into the original game content while the rescuer is training in the game. This will greatly increase the randomness of training and can be used to train unexpected emergencies, but it also increases the burden on instructors to create the scene. Flame-Sim \cite{43} developed by Flame Sim LLC is a commercial application training game that drives every firefighter to make decisions on the fire ground. It uses scenario generation technology, which allows the user to change or create a scene within a few minutes that can maximize the training time without a lot of set up time. Moreover, similar to Hazmat: Hotzone, Flame-Sim also can assist firefights in training many rescue operations such as rescue tool selection and use, room search, vents opened, and rescue people. Besides, Both games allow multiple participants to cooperate to complete tasks in a networked 3D game environment.
	
	In order to make the game environment more realistic, a large number of technologies and peripherals are used to simulate real disaster situations as possible, such as VR/AR/mixed reality(MR) technology, sensor technology, cave technology, and somatosensory technology. Virtual reality technology is the main way to enhance immersion and interactivity. Like the Nano Games sp. z o.o. \cite{44} company has designed a game allows simulate different types of traffic accidents. With the help of VR technology, rescuers are wearing protective clothing and equipped with real equipment can practice relevant rescue operations that provide appropriate assistance at the virtual accident site. Xu et al. \cite{45} have developed a VR-based game that can create an immersive environment to provide crane operators with an experiment opportunity to learn knowledge about how to deal with railway accidents. The FLAIM Trainer \cite{46} is a firefighting simulator that provides an immersive virtual environment that lets firefighters be trained in realistic emergency fire situations in a safe way than traditional training methods. It is not the first VR-based firefighting training simulator, but it is the first that combines virtual reality and haptic technologies. It use heated personal protective clothing and breathing apparatus to simulate all the sensory experiences that people might encounter in a real fire scenario, including extreme heat and difficulty breathing. At the same time, there are multiple scenarios in FLAIM Trainer which can train multiple base capabilities. Therefore, the immersion and other advantages brought by virtual reality technology provide a broad prospect for disaster rescue training. In addition to VR technology, Skovde University and the Swedish Rescue Services Agency further use CAVE (Cave Automatic Virtual Environment) technology to develop a game Sidh \cite{47} to train firefighters to wear respirators for search and rescue. During the game, participants can walk or run in a small room surrounded by screens, and move in a virtual environment to search for virtual victims. The movement speed in the virtual environment is controlled by the acceleration sensor installed on the player’s boots. Although the real-life situations simulated by virtual reality technology or CAVE technology has improved. It is still challenging to map the real world to the virtual world due to the limitations of current technology. Mixed reality simulation is an alternative way to balance between immersion and reality in training. Researches on the use of mixed reality games have shown the benefits of this approach in disaster response training. By using scenario-based mixed reality simulation, it provides support for team coordination and decision-making training, so that responders can coordinate face-to-face with each other in real time to simulate real disasters \cite{48}. The Icehouse game \cite{49} developed by the Lincoln Laboratory of the Massachusetts Institute of Technology provides a simulated game environment in which a group of disaster responders use a wearable computer and interface specifically designed for this game. In this live action game, players have to move in a physically simulated disaster space, which requires  physical exertion and team coordination to reduce virtual dangers and rescue virtual victims. Moreover, this game uses wearable technologies to monitor players’ information about their teammate and their own status (e.g. approximate distance between teammates, the distance from the leader, and heart rate). 
	
	Most of the previously mentioned games used to train rescue skills also support multiplayer mode (e.g. Hazmat: Hotzone, Flame-Sim, and Icehouse), and to a certain extent it can train the collaboration ability between rescuers. Moreover, there are some low-fidelity games/simulations that abstract the elements of cooperation in disasters specifically for training team cooperation. For example, The C3Fire \cite{50} simulation is a training environment that can train people’s team coordination awareness and team decision-making. It can generate a task environment in which participants cooperate to complete a specific mission, such as extinguishing a forest fire and save valuable houses. ZO Toups et al. \cite{51} developed a game called Team Coordination Game (TeC) which uses team coordination as the main component of the game's core mechanism. In this game distributed information among team members, participants need to gather information, cooperate, communicate, and rely on each other. 
	
	In addition, it is also very important to improve the moral judgment of rescuers. Because the rescuer often encounters various moral choices in the actual rescue process. If not handled properly, the rescuer may not only miss the best rescue opportunities, but also may develop mental illness like post-traumatic stress disorder (PTSD) \cite{52} \cite{53}. To solve this problem, some games are designed to train the ethical decision-making of inexperienced rescuers. Wahyudin et al. \cite{54} have designed a mobile first-person role-playing game (RPG) MAGNITUDE, which is used to train ethical decision-making in disaster situations. During the game, the player has to confront NPCs with ethical conflicts. For example, a boy’s leg was crushed by debris, and he lost a lot of blood. The heavy equipment used to clean up debris cannot be brought to the scene immediately. The player must choose to amputate or wait for equipment, both are very dangerous to the boy’s life. The player must take responsibility for the boy’s life.
	
	\begin{figure}[h]
		\centering
		\includegraphics[width=3.65in]{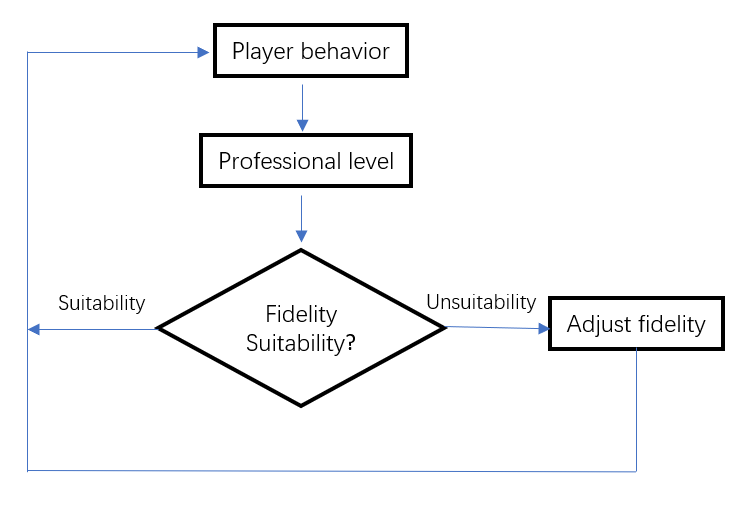}
		\caption{Game Fidelity Adaptive}
	\end{figure}

	With the development of science and technology, a large number of technologies continue to make the game training environment of disaster rescuers closer to the real-world. Meanwhile, most designers and developers of serious games usually strive for high-fidelity environments, especially in visual scenes. However, studies have shown that higher fidelity does not lead to higher learning efficiency. The relationship between fidelity and learning efficiency depends on the pre-knowledge and pre-skills that a trainee has mastered \cite{55}. In general, novices or intermediate trainees may be distracted by abundant irrelevant information and activities in the game which increase cognitive load, resulting in inefficient learning or training. Proficient knowledge and skills can make professional focus on the main purpose of the game, and high fidelity can bring them more immersion, so that high fidelity can bring professional more gain than others. At the same time, high fidelity brings more burdens of computing resources, causing a series of physical discomforts and lead to an increase in development costs. One way to overcome these challenges is to use low-fidelity simulation, which reduces the fidelity of the simulation and only focuses on concept learning. This low-fidelity simulation is economical where it can reduce the cost of the simulation whilst and increase the focus on the desired element of education \cite{51}. In order to prove the effectiveness of low-fidelity simulation, ZO Toups et al. \cite{51} tested 28 firefighters. The results showed that low fidelity is an effective training method when the demand for fidelity is not high. In summary, in the process of developing and designing such serious games, developers should design different games according to the professional level of the target group. For example,  novices can adopt a cartoon game style, whereas for professionals, the style of the game should be as realistic as possible. When the demand for fidelity is not high, or it is about training some specific knowledge, this knowledge can be abstracted develop a low-fidelity game. However, these solutions need to design a new game with different fidelity for people of different professional levels, which will greatly increase the development cost of the game. With the continuous development of artificial intelligence technology and adaptive technology, more and more games have begun to adjust game content based on the player’s behavior in the game. For example, in some answering games, the difficultly of questions will be appropriately adjusted according to the player’s correct rate of answering the question \cite{56}. Therefore, facing the relationship between game fidelity and learning efficiency, game fidelity adaption may be a feasible method, Fig 4. On the one hand, players’ data can be collected through their interaction with the game or by using some sensors (e.g. eye tracker, motion capturing, joysticks etc.) \cite{57} \cite{58}. Based on this data, algorithms such as decision trees, fuzzy logic, and others can be used to obtain the player’s professional level \cite{59}. On the other hand, the game fidelity is mainly expressed through the materials, textures, and models. Consequently, it may be possible to change the game material and model based on the player’s professional level in the game to achieve the purpose of fidelity adaptation. That will keep the relationship between the game environmental fidelity of the game and the professional level of player is maintained at the point of highest learning efficiency. We just provide a feasible idea for researchers to solve such problems. Thus, the specific implementation of the game fidelity adaptive method requires further research and exploration.

	\subsubsection{SGDRs for Commanders}
	once an emergency occurs, incident commanders of rescue service will rapidly assess the situation before deploying their resources. This evaluation and command are extremely important to achieve the successful result of the rescue at this stage. At this point, improper decision-making or poorly informed can cause catastrophic consequences, leading to loss of life or avoidable property loss. In order to manage this situation, commanders always have to make the correct decision even when they are under the extreme pressures of the incident. In an attempt to provide a realistic training environment, some games combine videos, photos, and role-playing to recreate the scene \cite{60}. However, videos and photos will limit the flexibility of the scene and cannot be used repeatedly. Therefore, many games based on the random event have been developed which. In these games, some emergencies are randomly generated, and players need to resolve these events reasonably according to the actual situation.
	
	Most of the games are about a micromanaging emergency where the player can directly control a selection of rescue units (e.g. firefighter, medical staff, and police, etc.) to save lives as possible. These games are often used by incident commanders to train their ability to collect information and allocate disaster rescue resources. For example, Emergency \cite{61} is a disaster strategy game developed by Sixteen Tons Entertainment. In Emergency, the player has a city map that points out the location of various rescue resources, and the place where the emergency occurs. The player needs to mobilize limited resources to resolve the emergency. This is a complicated process in which the player needs to observe what is happening at the emergency location, then dispatches the needed rescue resources and vehicles. For example, the player must first mobilize ambulance or helicopter when seeing blood, handle large-scale fires which require heavy fire trucks with water guns, and need to control firefighters to cut down trees to create fire barriers when dealing with the forest fires. Another game is KriseIM \cite{62}, In this game, players will get information from the notification window, and then respond to the crisis by dispatching police, ambulances and fire departments. At end of the event, the player and instructor will be provided a debriefing to show the results of the training. Rescue: Everyday Heroes \cite{63} focuses only on the fire. In this game, the player keeps track of several fire stations and has to come up with a strategy for the mission and chooses the most suitable fire extinguishing agent according to different circumstances. In these games, artificial intelligence technology is used to control a large number of NPCs’ behavior. Especially those NPCs affected by the disaster. At the same time, the player needs to control virtual teammates as the response team members to rescue the NPCs affected by the disaster. However, those virtual teammates will blindly follow the player's command and take action, regardless of whether these commands are reasonable. In order to make the commander consider the safety and emotions of the teammate during training. Djordjevich et al. \cite{64} developed an emotional model into virtual teammates for a training game Ground Truth: Toxic City. This game allows players to play the role of the incident commander, command virtual teammates to deal with the leakage of toxic chemicals,aim to training the commander's strategy. In this game, Artificial neural network computations are used to model intensities of fear and anger and then convert them into fuzzy sets to be easier to understand by the human cognitive model. If the player orders the virtual teammate to take inappropriate actions (e.g. require going deep inside the hazardous fog), the virtual teammate’s emotional state may change dynamically. This will cause the virtual teammates to disobey the command or reduce its ability to perform other requested actions. Integrating emotions into the virtual teammates will prompt the player to consider the rationality of instructions and the emotional state of the members in the process of commanding rescue. This will more accurately reflect the player’s decisions in such situations.
	
	After a disaster occurs, it is often necessary for the incident commander to make a Rapid Damage Assessment (R.D.A.). Sooraj K Babu et al. \cite{65} came up with a multiplayer game, which is set in an earthquake. Each player has to take the role of head of a department to deal with the rescue operation. Players must first take a R.D.A. according to destroyed buildings and injured people. Then each player collaborates to provide resources and assistance according to the R.D.A level and the responsibilities of the department they played. In addition, many studies have shown that the main way for incident commanders to know about the disaster is through social media \cite{66} \cite{67}. The incident commander should not only be trained to obtain information from the disaster site for damage assessment, but also need to learn to extract useful information from social media. Abbasi et al. \cite{68} designed a live action game in which the player can get information from social media (e.g. Internet, TV, newspaper, E-mail, telephone, etc.). Multiple players must exchange information to determine the level and situation of the disaster in which commanders can be trained to collect information from media and communication ability.
	
	In the actual rescue process, the commander often needs to cooperate with rescuers on the scene of the disaster to complete the task. In this way, real-time communication and cooperation between commanders and rescuers are very important. However, for such serious games, the training of cooperation factors often concentrated on the cooperation between commanders of different departments. Few games involve commanding cooperation between commanders and rescuers in which will provide developers with another design solution.

	\subsection{SGDRs for Recovery Stage}
	When a disaster is over, building back homes is very necessary. This stage usually occurs six months after the disaster and lasts for long time. In the recovery phase, the relief work is no longer just to provide some emergency supplies and search or rescue the victims, but it is rather a series of longer-term assistance, including returning people affected by the disaster to normal life and strengthen future anti-disaster capacity. However, post-disaster reconstruction will be limited by many factors such as regional culture, funding, and the extent of the disaster. Therefore, serious games are not suitable for training in this aspect. Only a few games incorporate the element of improving environment and living conditions of communities. For example, Hazagora \cite{23} is a board game in which players represent the inhabitants of a volcanic, who must develop and maintain communities. After the disaster, the player needs to take some measures to maintain communities, such as removing the destroyed buildings, burying the dead, and clean up contaminated resources (e.g. food, natural water). In this way, the player can experience the impact of the disaster and learn post-disaster strategies to mitigate the continuing impact after the disaster. 
	
	\section{Discussion and Prospects}
	Disasters occur every day in the world that affects thousands of people. In order to reduce the losses caused by disasters, participants in disaster relief must perform their duties and train themselves to prepare for disaster. Many studies have proved that serious games are a good way to train relevant participants. However, SGDRs are not exhaustive and has a series of limitations. Considering these deficiencies, the following aspects of SGDRs are worth to be further attention.
	
	The content of SGDRs are not comprehensive. On the one hand, SGDRs could not consider regional cultural diversity and custom-imposed taboos or local needs, which will limit its use range. On the other hand, most SGDRs focus only on most common disasters like fire, earthquake, floods, and tsunami. A few games involve other disasters like droughts, extreme weather, and disease epidemics. Therefore, developers should take this aspect into consideration when developing SGDRs in the future, and continuously enrich their content.
	
	Game feedback, including in-game feedback and post-game feedback, is particularly important for SGDRs \cite{69}, but it is often ignored in many SGDRs. The in-game feedback can judge the player’s in-game operations, affirm the correct operations, and punish the wrong operations, if there is no proper feedback in the game, the player can make mistakes at will in the game so that these mistakes can be ignored in real-world operations and lead to disastrous consequences. The post-game mainly is debriefing which can offer an opportunity for the player to process and consolidate their in-game operation. Therefore, when the player completes an operation in SGDRs, feedback should be given to the player in certain forms (e.g. sound, animation, special effects, etc.) based on the player’s operation, and the player should be given a debriefing when the game is over.
	
	Some articles indicate that game control complexity and game environment can affect SGDRs training \cite{31}. That requires the simple operation of SGDRs, and players should be familiar with the game environment before training. At the same time, developers should design different SGDRs according to the player’s professional level. novices can be trained by simply quiz games. The professional can be trained through processes or problem-solving in a complex game environment.
	
	For evaluation of the effectiveness of SGDRs, there are few detailed descriptions of comprehensive evaluation models, which leave room for future research and speculations. Only a few games have been tested using player feedback. Their research results show that serious games can be effectively used in simulations, training many disaster relief related activities and increasing disaster awareness. However, to face of a lack of professional evaluation, we recommend using a combination of player feedback and professional opinions to evaluate the serious game in the realm of disaster relief.
	
	\section{Conclusion}
	SGDRs is an effective method for disaster relief training and being intensively studied. In order to show the effectiveness of serious games for disaster relief training, this paper investigated the traditional methods and serious game methods for disaster relief training and determined that the use of serious game training can make up for the limitations of the traditional methods. Apart from that, due to the absence of systematic description of disaster relief work in different stages of disaster relief, we introduced disaster management and divided disaster relief into three stages: Preparedness, Response, and Recovery. Then, based on the different stages of disaster relief, the technologies and functions of SGDRs were summarized and analyzed. Finally, we discussed the current deficiencies of SGDRs. To sum up, our work can provide a guidance for participants in disaster relief work and training. Meanwhile, we provide suggestions for researchers to design more effective serious games for disaster relief.


	\ifCLASSOPTIONcaptionsoff
	\newpage
	\fi
	
	
	
	\bibliographystyle{IEEEtran}
	\bibliography{A Review on Serious Games for Disaster Relief}

\begin{thebibliography}{10}
\providecommand{\url}[1]{#1}
\csname url@samestyle\endcsname
\providecommand{\newblock}{\relax}
\providecommand{\bibinfo}[2]{#2}
\providecommand{\BIBentrySTDinterwordspacing}{\spaceskip=0pt\relax}
\providecommand{\BIBentryALTinterwordstretchfactor}{4}
\providecommand{\BIBentryALTinterwordspacing}{\spaceskip=\fontdimen2\font plus
\BIBentryALTinterwordstretchfactor\fontdimen3\font minus
  \fontdimen4\font\relax}
\providecommand{\BIBforeignlanguage}[2]{{%
\expandafter\ifx\csname l@#1\endcsname\relax
\typeout{** WARNING: IEEEtran.bst: No hyphenation pattern has been}%
\typeout{** loaded for the language `#1'. Using the pattern for}%
\typeout{** the default language instead.}%
\else
\language=\csname l@#1\endcsname
\fi
#2}}
\providecommand{\BIBdecl}{\relax}
\BIBdecl

\bibitem{1}
J.~F. Waeckerle, S.~R. Lillibridge, F.~M. Burkle~Jr, and E.~K. Noji, ``Disaster
  medicine: challenges for today,'' \emph{Annals of emergency medicine},
  vol.~23, no.~4, pp. 715--718, 1994.

\bibitem{2}
R.~Below and P.~Wallemacq, ``Natural disasters 2017,'' \emph{Brussels: Centre
  for Research on the Epidemiology of Disasters (CRED)}, 2018.

\bibitem{3}
I.~M. Shaluf, ``An overview on disasters,'' \emph{Disaster Prevention and
  Management: An International Journal}, 2007.

\bibitem{4}
K.~Meesters, L.~Olthof, and B.~Van~de Walle, ``Disaster in my backyard: A
  serious game to improve community disaster resilience,'' in \emph{Proceedings
  of the European Conference on Games Based Learning}, vol.~2, 2014, pp.
  714--722.

\bibitem{5}
T.~M. Connolly, E.~A. Boyle, E.~MacArthur, T.~Hainey, and J.~M. Boyle, ``A
  systematic literature review of empirical evidence on computer games and
  serious games,'' \emph{Computers \& education}, vol.~59, no.~2, pp. 661--686,
  2012.

\bibitem{6}
W.~N. Carter, ``Disaster management: A disaster manager's handbook,'' 2008.

\bibitem{7}
``Disaster management,''
  \url{https://www.ifrc.org/en/what-we-do/disaster-management/about-disaster-management/},
  2020.

\bibitem{8}
M.~J. Sim{\~o}es-Marques, ``Facing disasters—trends in applications to
  support disaster management,'' in \emph{Advances in Human Factors and System
  Interactions}.\hskip 1em plus 0.5em minus 0.4em\relax Springer, 2017, pp.
  203--215.

\bibitem{9}
D.~Paton, ``Disaster relief work: An assessment of training effectiveness,''
  \emph{Journal of Traumatic stress}, vol.~7, no.~2, pp. 275--288, 1994.

\bibitem{10}
A.~Malizia and .~C.~M. Group, ``Disaster management in case of cbrne events: an
  innovative methodology to improve the safety knowledge of advisors and first
  responders,'' \emph{Defense \& Security Analysis}, vol.~32, no.~1, pp.
  79--90, 2016.

\bibitem{11}
G.~B. Green, S.~Modi, K.~Lunney, and T.~L. Thomas, ``Generic evaluation methods
  for disaster drills in developing countries,'' \emph{Annals of emergency
  medicine}, vol.~41, no.~5, pp. 689--699, 2003.

\bibitem{12}
K.~Meesters and B.~Van~de Walle, ``Disaster in my backyard: A serious game
  introduction to disaster information management.'' in \emph{ISCRAM}, 2013.

\bibitem{13}
X.~Yang, Z.~Wu, and Y.~Li, ``Difference between real-life escape panic and
  mimic exercises in simulated situation with implications to the statistical
  physics models of emergency evacuation: The 2008 wenchuan earthquake,''
  \emph{Physica A: Statistical Mechanics and its Applications}, vol. 390,
  no.~12, pp. 2375--2380, 2011.

\bibitem{14}
J.~B. Staal, J.~Rainville, J.~Fritz, W.~Van~Mechelen, and G.~Pransky,
  ``Physical exercise interventions to improve disability and return to work in
  low back pain: current insights and opportunities for improvement,''
  \emph{Journal of occupational Rehabilitation}, vol.~15, no.~4, pp. 491--505,
  2005.

\bibitem{15}
C.~A. Evans, M.~Baumberger-Henry, R.~Schwartz, and T.~Veenema, ``Nursing
  students' transfer of learning during a disaster tabletop exercise,''
  \emph{Nurse educator}, vol.~44, no.~5, pp. 278--283, 2019.

\bibitem{16}
J.~L. Taylor, B.~J. Roup, D.~Blythe, G.~K. Reed, T.~A. Tate, and K.~A. Moore,
  ``Pandemic influenza preparedness in maryland: improving readiness through a
  tabletop exercise,'' \emph{Biosecurity and bioterrorism: biodefense strategy,
  practice, and science}, vol.~3, no.~1, pp. 61--69, 2005.

\bibitem{17}
S.~YOUSEFI, H.~Khankeh, Y.~Akbari, A.~Dalvandi, and E.~Bakhshi, ``The effect of
  the implementation of the national program for hospital preparedness on the
  readiness of nurses under simulated conditions of incidents and disasters,''
  2016.

\bibitem{18}
A.~M. Wendelboe, J.~Amanda~Miller, D.~Drevets, L.~Salinas, E.~Miller,
  D.~Jackson, A.~Chou, J.~Jill~Raines, P.~H.~W. Group \emph{et~al.}, ``Tabletop
  exercise to prepare institutions of higher education for an outbreak of
  covid-19,'' \emph{Journal of Emergency Management}, vol.~18, no.~2, pp.
  1--20, 2020.

\bibitem{19}
``What is a tabletop exercise,''
  \url{https://uwpd.wisc.edu/content/uploads/2014/01/What_is_a_tabletop_exercise.pdf},
  2020.

\bibitem{20}
T.~Susi, M.~Johannesson, and P.~Backlund, ``Serious games: An overview,'' 2007.

\bibitem{21}
I.~Heldal and C.~H. Wijkmark, ``Simulations and serious games for firefighter
  training: Users' perspective.'' in \emph{ISCRAM}, 2017.

\bibitem{22}
S.~Sharples, S.~Cobb, A.~Moody, and J.~R. Wilson, ``Virtual reality induced
  symptoms and effects (vrise): Comparison of head mounted display (hmd),
  desktop and projection display systems,'' \emph{Displays}, vol.~29, no.~2,
  pp. 58--69, 2008.

\bibitem{23}
S.~Mossoux, A.~Delcamp, S.~Poppe, C.~Michellier, F.~Canters, and M.~Kervyn,
  ``Hazagora: will you survive the next disaster?--a serious game to raise
  awareness about geohazards and disaster risk reduction.'' \emph{Natural
  Hazards \& Earth System Sciences}, vol.~16, no.~1, 2016.

\bibitem{24}
Z.~Liu, A.~Kashef, G.~Lougheed, and N.~Benichou, ``Review of three dimensional
  water fog techniques,'' 2002.

\bibitem{25}
``Emergency management,''
  \url{https://en.wikipedia.org/wiki/Emergency_management}, 2020.

\bibitem{26}
S.~J. Mannsverk, I.~Di~Loreto, and M.~Divitini, ``Flooded: a location-based
  game for promoting citizens’ preparedness to flooding situations,'' in
  \emph{International Conference on Games and Learning Alliance}.\hskip 1em
  plus 0.5em minus 0.4em\relax Springer, 2013, pp. 90--103.

\bibitem{27}
``Disaster master,''
  \url{https://www.preventionweb.net/educational/view/60696}, 2020.

\bibitem{28}
I.~Kerlow, ``Earth girl saves the day: a computer game prototype about earth
  hazards,'' in \emph{2011 Second International Conference on Culture and
  Computing}.\hskip 1em plus 0.5em minus 0.4em\relax IEEE, 2011, pp. 33--38.

\bibitem{29}
G.~Jacob, R.~Jayakrishnan, and K.~Bijlani, ``Smart fire safety: Serious game
  for fire safety awareness,'' in \emph{Information and Decision
  Sciences}.\hskip 1em plus 0.5em minus 0.4em\relax Springer, 2018, pp. 39--47.

\bibitem{30}
A.~Rahouti, G.~Salze, R.~Lovreglio, and S.~Datoussa{\"\i}d, ``An immersive
  serious game for firefighting and evacuation training in healthcare
  facilities,'' \emph{Int. J. Comput. Electr. Autom. Control Inf. Eng},
  vol.~11, pp. 1038--1044, 2017.

\bibitem{31}
J.~Ribeiro, J.~E. Almeida, R.~J. Rossetti, A.~Coelho, and A.~L. Coelho, ``Using
  serious games to train evacuation behaviour,'' in \emph{7th Iberian
  Conference on Information Systems and Technologies (CISTI 2012)}.\hskip 1em
  plus 0.5em minus 0.4em\relax IEEE, 2012, pp. 1--6.

\bibitem{32}
P.~A. Ruffino, D.~Permadi, M.~B. Mahadzir, A.~Osello, and A.~B. Aris,
  ``Simulation and serious game for fire evacuation training.''

\bibitem{33}
U.~R{\"u}ppel and K.~Schatz, ``Designing a bim-based serious game for fire
  safety evacuation simulations,'' \emph{Advanced engineering informatics},
  vol.~25, no.~4, pp. 600--611, 2011.

\bibitem{34}
H.~Liang, F.~Liang, F.~Wu, C.~Wang, and J.~Chang, ``Development of a vr
  prototype for enhancing earthquake evacuee safety,'' in \emph{Proceedings of
  the 16th ACM SIGGRAPH International Conference on Virtual-Reality Continuum
  and its Applications in Industry}, 2018, pp. 1--8.

\bibitem{35}
R.~Lovreglio and M.~Kinateder, ``Augmented reality for pedestrian evacuation
  research: promises and limitations,'' \emph{Safety science}, vol. 128, p.
  104750, 2020.

\bibitem{36}
C.~Catal, A.~Akbulut, B.~Tunali, E.~Ulug, and E.~Ozturk, ``Evaluation of
  augmented reality technology for the design of an evacuation training game,''
  \emph{Virtual Reality}, pp. 1--10, 2019.

\bibitem{37}
G.~Rebolledo-Mendez, K.~Avramides, S.~de~Freitas, and K.~Memarzia, ``Societal
  impact of a serious game on raising public awareness: the case of floodsim,''
  in \emph{Proceedings of the 2009 ACM SIGGRAPH symposium on video games},
  2009, pp. 15--22.

\bibitem{38}
G.~Pereira, R.~Prada, and A.~Paiva, ``Disaster prevention social awareness: The
  stop disasters! case study,'' in \emph{2014 6th International Conference on
  Games and Virtual Worlds for Serious Applications (VS-GAMES)}.\hskip 1em plus
  0.5em minus 0.4em\relax IEEE, 2014, pp. 1--8.

\bibitem{39}
``Buzz about dengue,''
  \url{https://www.climatecentre.org/downloads/modules/games/A%20Buzz%20about%20Dengue%20.pdf},
  2020.

\bibitem{40}
``Dissolving disasters,''
  \url{https://www.climatecentre.org/resources-games/games/5/dissolving-disasters},
  2020.

\bibitem{41}
S.~M. Walton, K.~M. Conrad, S.~E. Furner, and D.~G. Samo, ``Cause, type, and
  workers' compensation costs of injury to fire fighters,'' \emph{American
  journal of industrial medicine}, vol.~43, no.~4, pp. 454--458, 2003.

\bibitem{42}
``Hazmat: Hotzone,'' \url{https://www.etc.cmu.edu/projects/hazmat_2005/}, 2020.

\bibitem{43}
K.~Hullett and M.~Mateas, ``Scenario generation for emergency rescue training
  games,'' in \emph{Proceedings of the 4th International Conference on
  Foundations of Digital Games}, 2009, pp. 99--106.

\bibitem{44}
J.~K. Argasi{\'n}ski, P.~W{\k{e}}grzyn, and P.~Strojny, ``Affective vr serious
  game for firefighter training,'' in \emph{Workshop on affective computing and
  context awareness in ambient intelligence}, vol.~41, 2018, p.~43.

\bibitem{45}
J.~Xu, Z.~Tang, X.~Yuan, Y.~Nie, Z.~Ma, X.~Wei, and J.~Zhang, ``A vr-based the
  emergency rescue training system of railway accident,'' \emph{Entertainment
  Computing}, vol.~27, pp. 23--31, 2018.

\bibitem{46}
M.~Vackov{\'a}, P.~Lo{\v{s}}onczi, J.~Drot{\'a}rov{\'a}, and
  L.~Kov{\'a}{\v{c}}ov{\'a}, ``The use of virtual reality resources to increase
  safety in the training of fire and rescue corps units.''

\bibitem{47}
P.~Backlund, H.~Engstrom, C.~Hammar, M.~Johannesson, and M.~Lebram, ``Sidh--a
  game based firefighter training simulation,'' in \emph{2007 11th
  International Conference Information Visualization (IV'07)}.\hskip 1em plus
  0.5em minus 0.4em\relax IEEE, 2007, pp. 899--907.

\bibitem{48}
J.~E. Fischer, W.~Jiang, A.~Kerne, C.~Greenhalgh, S.~D. Ramchurn, S.~Reece,
  N.~Pantidi, and T.~Rodden, ``Supporting team coordination on the ground:
  requirements from a mixed reality game,'' in \emph{COOP 2014-Proceedings of
  the 11th International Conference on the Design of Cooperative Systems, 27-30
  May 2014, Nice (France)}.\hskip 1em plus 0.5em minus 0.4em\relax Springer,
  2014, pp. 49--67.

\bibitem{49}
S.~A. Alharthi, H.~N. Sharma, S.~Sunka, I.~Dolgov, and Z.~O. Toups, ``Designing
  future disaster response team wearables from a grounding in practice,'' in
  \emph{Proceedings of the Technology, Mind, and Society}, 2018, pp. 1--6.

\bibitem{50}
R.~Granlund, B.~Johansson, and M.~Persson, ``C3fire: A microworld for
  collaboration training in the rolf environment,'' in \emph{42nd Conference on
  Simulation and Modelling, Simulation in Theory and Practice, 8-9 October
  2001, Porsgrunn, Norway}, 2001.

\bibitem{51}
Z.~O. Toups, A.~Kerne, and W.~A. Hamilton, ``The team coordination game:
  Zero-fidelity simulation abstracted from fire emergency response practice,''
  \emph{ACM Transactions on Computer-Human Interaction (TOCHI)}, vol.~18,
  no.~4, pp. 1--37, 2011.

\bibitem{52}
W.~Berger, E.~S.~F. Coutinho, I.~Figueira, C.~Marques-Portella, M.~P. Luz,
  T.~C. Neylan, C.~R. Marmar, and M.~V. Mendlowicz, ``Rescuers at risk: a
  systematic review and meta-regression analysis of the worldwide current
  prevalence and correlates of ptsd in rescue workers,'' \emph{Social
  psychiatry and psychiatric epidemiology}, vol.~47, no.~6, pp. 1001--1011,
  2012.

\bibitem{53}
C.~Holmg{\aa}rd, G.~N. Yannakakis, K.-I. Karstoft, and H.~S. Andersen, ``Stress
  detection for ptsd via the startlemart game,'' in \emph{2013 Humaine
  Association Conference on Affective Computing and Intelligent
  Interaction}.\hskip 1em plus 0.5em minus 0.4em\relax IEEE, 2013, pp.
  523--528.

\bibitem{54}
D.~Wahyudin and S.~Hasegawa, ``Mobile serious game design for training ethical
  decision making skills of inexperienced disaster volunteers,'' \emph{The
  Journal of Information and Systems in Education}, vol.~14, no.~1, pp. 28--41,
  2015.

\bibitem{55}
X.~Ye, P.~Backlund, J.~Ding, and H.~Ning, ``Fidelity in simulation-based
  serious games,'' \emph{IEEE Transactions on Learning Technologies}, 2019.

\bibitem{56}
S.~Klinkenberg, M.~Straatemeier, and H.~L. van~der Maas, ``Computer adaptive
  practice of maths ability using a new item response model for on the fly
  ability and difficulty estimation,'' \emph{Computers \& Education}, vol.~57,
  no.~2, pp. 1813--1824, 2011.

\bibitem{57}
L.~Zhang, J.~W. Wade, D.~Bian, A.~Swanson, Z.~Warren, and N.~Sarkar, ``Data
  fusion for difficulty adjustment in an adaptive virtual reality game system
  for autism intervention,'' in \emph{International Conference on
  Human-Computer Interaction}.\hskip 1em plus 0.5em minus 0.4em\relax Springer,
  2014, pp. 648--652.

\bibitem{58}
D.~Puzenat and I.~Verlut, ``Behavior analysis through games using artificial
  neural networks,'' in \emph{2010 Third International Conference on Advances
  in Computer-Human Interactions}.\hskip 1em plus 0.5em minus 0.4em\relax IEEE,
  2010, pp. 134--138.

\bibitem{59}
M.~Frutos-Pascual and B.~G. Zapirain, ``Review of the use of ai techniques in
  serious games: Decision making and machine learning,'' \emph{IEEE
  Transactions on Computational Intelligence and AI in Games}, vol.~9, no.~2,
  pp. 133--152, 2015.

\bibitem{60}
W.~Viant, J.~Purdy, and J.~Wood, ``Serious games for fire and rescue
  training,'' in \emph{2016 8th Computer Science and Electronic Engineering
  (CEEC)}.\hskip 1em plus 0.5em minus 0.4em\relax IEEE, 2016, pp. 136--139.

\bibitem{61}
I.~Di~Loreto, S.~Mora, and M.~Divitini, ``Collaborative serious games for
  crisis management: an overview,'' in \emph{2012 IEEE 21st International
  Workshop on Enabling Technologies: Infrastructure for Collaborative
  Enterprises}.\hskip 1em plus 0.5em minus 0.4em\relax IEEE, 2012, pp.
  352--357.

\bibitem{62}
S.~E. Skogen and J.~Radianti, ``Architecture of high fidelity simulation tool
  for crisis management training,'' in \emph{2018 5th International Conference
  on Information and Communication Technologies for Disaster Management
  (ICT-DM)}.\hskip 1em plus 0.5em minus 0.4em\relax IEEE, 2018, pp. 1--4.

\bibitem{63}
``Rescue: Everyday heroes,''
  \url{https://www.metacritic.com/game/pc/rescue-2-everyday-heroes}, 2020.

\bibitem{64}
D.~D. Djordjevich, P.~G. Xavier, M.~L. Bernard, J.~H. Whetzel, M.~R. Glickman,
  and S.~J. Verzi, ``Preparing for the aftermath: Using emotional agents in
  game-based training for disaster response,'' in \emph{2008 IEEE Symposium On
  Computational Intelligence and Games}.\hskip 1em plus 0.5em minus 0.4em\relax
  IEEE, 2008, pp. 266--275.

\bibitem{65}
S.~K. Babu, M.~L. McLain, K.~Bijlani, R.~Jayakrishnan, and R.~R. Bhavani,
  ``Collaborative game based learning of post-disaster management: Serious game
  on incident management frameworks for post disaster management,'' in
  \emph{2016 IEEE Eighth International Conference on Technology for Education
  (T4E)}.\hskip 1em plus 0.5em minus 0.4em\relax IEEE, 2016, pp. 80--87.

\bibitem{66}
H.~Gao, G.~Barbier, and R.~Goolsby, ``Harnessing the crowdsourcing power of
  social media for disaster relief,'' \emph{IEEE Intelligent Systems}, vol.~26,
  no.~3, pp. 10--14, 2011.

\bibitem{67}
S.~Kumar, G.~Barbier, M.~A. Abbasi, and H.~Liu, ``Tweettracker: An analysis
  tool for humanitarian and disaster relief.'' \emph{ICwSM}, vol.~11, pp.
  78--82, 2011.

\bibitem{68}
M.-A. Abbasi, S.~Kumar, J.~A. Andrade~Filho, and H.~Liu, ``Lessons learned in
  using social media for disaster relief-asu crisis response game,'' in
  \emph{International Conference on Social Computing, Behavioral-Cultural
  Modeling, and Prediction}.\hskip 1em plus 0.5em minus 0.4em\relax Springer,
  2012, pp. 282--289.

\bibitem{69}
W.~S. Ravyse, A.~S. Blignaut, V.~Leendertz, and A.~Woolner, ``Success factors
  for serious games to enhance learning: a systematic review,'' \emph{Virtual
  Reality}, vol.~21, no.~1, pp. 31--58, 2017.

\end{thebibliography}
	
	%
	%
	%
	
	%


	\begin{IEEEbiography}[{\includegraphics[width=1in,height=1.25in,clip,keepaspectratio]{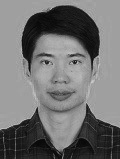}}]{\textbf{Huansheng Ning}}
	received his B.S. degree from Anhui University in 1996 and his Ph.D. degree from Beihang University in 2001. He is currently a Professor and Vice Dean with the School of Computer and Communication Engineering, University of Science and Technology Beijing and China and Beijing Engineering Research Center for Cyberspace Data Analysis and Applications, China, and the founder and principal at Cybermatics and Cyberspace International Science and Technology Cooperation Base. He has authored several books and over 70 papers in journals and at international conferences/workshops. He has been the Associate Editor of IEEE Systems Journal and IEEE Internet of Things Journal, Chairman (2012) and Executive Chairman (2013) of the program committee at the IEEE international Internet of Things conference, and the Co-Executive Chairman of the 2013 International cyber technology conference and the 2015 Smart World Congress. His awards include the IEEE Computer Society Meritorious Service Award and the IEEE Computer Society Golden Core Member Award. His current research interests include Internet of Things, Cyber Physical Social Systems, electromagnetic sensing and computing.
	\end{IEEEbiography}
	
	\begin{IEEEbiography}[{\includegraphics[width=1in,height=1.25in,clip,keepaspectratio]{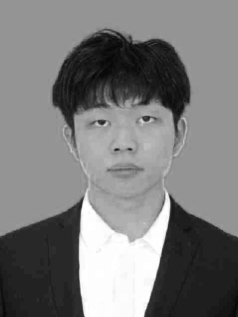}}]{Zhangfeng Pi}
		received his B.S. degree from Tianjin College of University of Science and Technology Beijing and currently is working toward his M.S. degree in the School of Computer and Communication Engineering, University of Science and Technology Beijing, China. His current research is about serious games.
	\end{IEEEbiography}
	
	\begin{IEEEbiography}[{\includegraphics[width=1in,height=1.25in,clip,keepaspectratio]{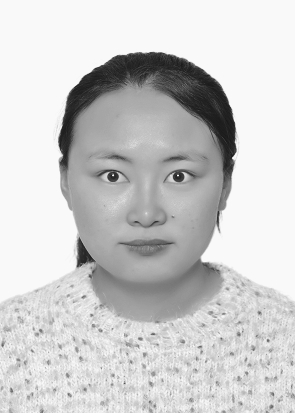}}]{Wenxi Wang}
		received her B.E. degree from Ludong University in 2019. She is currently pursuing her PhD. degree in the School of Computer and Communication Engineering, University of Science and Technology Beijing. Her current research interests include Serious Games and Social Computing.
	\end{IEEEbiography}
	
	\begin{IEEEbiography}[{\includegraphics[width=1in,height=1.25in,clip,keepaspectratio]{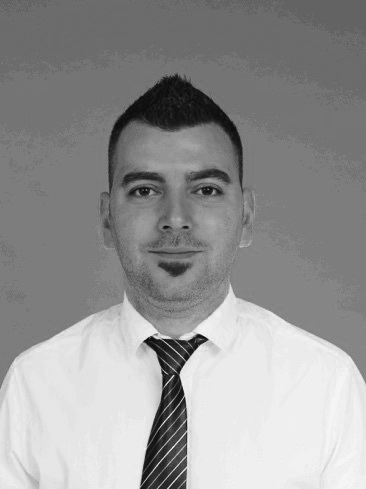}}]{Fadi Farha}
		received his MS degree and currently work- ing toward Ph.D. degree in the School of Computer and Communication Engineering, University of Science and Technology Beijing, China. His current research interests include Physical Unclonable Function (PUF), Smart Home, Security Solutions, ZigBee, Computer Architecture and Hardware Security.
	\end{IEEEbiography}
	
	\begin{IEEEbiography}[{\includegraphics[width=1in,height=1.25in,clip,keepaspectratio]{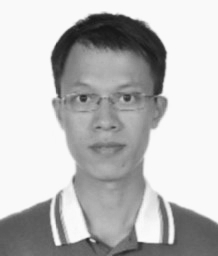}}]{Shunkun Yang}
		received his B.S., M.S., and Ph.D. degrees from the School of Reliability and Systems Engineering at Beihang University in 2000, 2003, and 2011, respectively. He is an associate research professor with Beihang University since 2016. He is also an associate research scientist of Columbia University through 2014.09 to 2015.09. His main research interests are reliability, testing and diagnosis for embedded software, CPS, IoT, Intelligent manufacturing, etc.
	\end{IEEEbiography}


	
	

\end{document}